\documentclass[aip,rsi,amsmath,amssymb,graphicx,reprint]{revtex4-1}
\usepackage{leftidx}
\usepackage{graphicx}
\usepackage[colorlinks=true, pdfstartview=FitV, linkcolor=blue, citecolor=blue, urlcolor=blue]{hyperref}
\usepackage[subrefformat=parens,labelformat=parens,position=t,caption=false]{subfig}
\usepackage{dcolumn}
\usepackage{bm}

\begin{document}

\preprint{RSI/123-QED}
\title{Active Stabilization of a Diode Laser Injection Lock} 

\author{Brendan Saxberg}
\affiliation{Department of Physics, University of Washington, P.O. Box 351560, Seattle, Washington 98195-1560, USA}
\author{Benjamin Plotkin-Swing}
\affiliation{Department of Physics, University of Washington, P.O. Box 351560, Seattle, Washington 98195-1560, USA}
\author{Subhadeep Gupta}
\affiliation{Department of Physics, University of Washington, P.O. Box 351560, Seattle, Washington 98195-1560, USA}

\date{\today}

\begin{abstract} We report on a device to electronically stabilize the optical injection lock of a semiconductor diode laser. Our technique uses as discriminator the peak height of the laser's transmission signal on a scanning Fabry-Perot cavity and feeds back to the diode current, thereby maintaining maximum optical power in the injected mode. A two-component feedback algorithm provides constant optimization of the injection lock, keeping it robust to slow thermal drifts and allowing fast recovery from sudden failures such as temporary occlusion of the injection beam. We demonstrate the successful performance of our stabilization method in a diode laser setup at 399 nm used for laser cooling of Yb atoms. The device eases the requirements on passive stabilization and can benefit any diode laser injection lock application, particularly those where several such locks are employed.
\end{abstract}


\maketitle 


\section{\label{sec:Introduction}Introduction}

Narrow linewidth semiconductor diode lasers have myriad applications in high-resolution spectroscopy \cite{tanner,gustavsson,Lenth} with great popularity in the field of atomic, molecular and optical (AMO) physics where narrow atomic transitions can be suitably addressed for a large variety of wavelengths \cite{wieman}. A common method to simultaneously achieve high power and narrow linewidth is to transfer the narrow frequency spectrum of a well-stabilized but low-power ``master'' laser to a high-power broad-spectrum ``slave'' diode laser by means of optical injection \cite{hadley,kozuma,lucas,dualspecies,fourinjlockthesis,fourinjlockpaper}. Such injection locks are widely used in experimental setups for laser cooling and trapping of atoms \cite{kozuma,dualspecies,fourinjlockthesis,fourinjlockpaper,yabuzaki}, and typically rely on passive stability through thermal isolation, high injection light power, and mechanical isolation \cite{kozuma}. They are thus very sensitive to ambient conditions and may require frequent user interventions. Experiments that need multiple injection locks to work simultaneously, such as those with limited master laser power \cite{yabuzaki} or experiments with multiple atomic species \cite{dualspecies,fourinjlockthesis,fourinjlockpaper}, are particularly vulnerable to the failure of any one injection lock. 

In this paper we describe an active stabilization technique for a diode laser injection lock which preserves the desired narrow spectrum against ambient thermal and mechanical fluctuations. We utilize a microcontroller which adds an offset to the slave diode current based on the slave spectrum measured in a Fabry-Perot interferometer. We apply this setup to a $399\,$nm diode laser injection lock, suitable for the $\leftidx{^1}{S}{_0} \rightarrow \leftidx{^1}{P}{_1}$ transition in ytterbium (Yb) used in laser cooling applications. We demonstrate the ability to maintain an injection lock while the slave diode is undergoing a change in thermal conditions, when the standard injection lock is unstable. Our device reduces the requirements for passive stabilization against thermal and mechanical disturbances, and allows stable operation at low values of seed (injection) power. The stabilization technique is quite general and is also applicable to diode laser setups at other wavelengths. 

\section{\label{sec:Setup}Experimental Setup}

\begin{figure}
        \centering
        \subfloat[]{
                \includegraphics{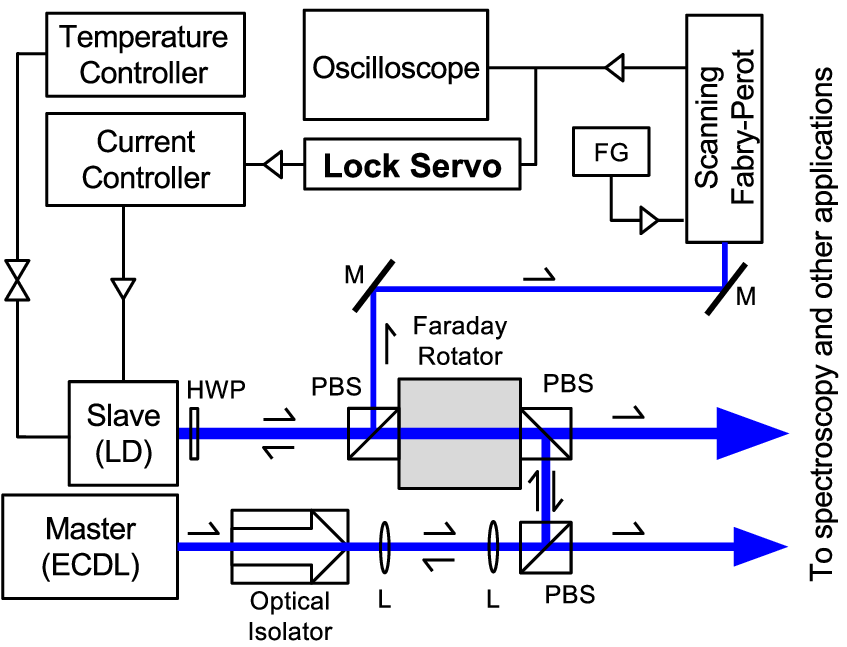}
                \label{fig:diagram}
                }

        \subfloat[]{
                \includegraphics{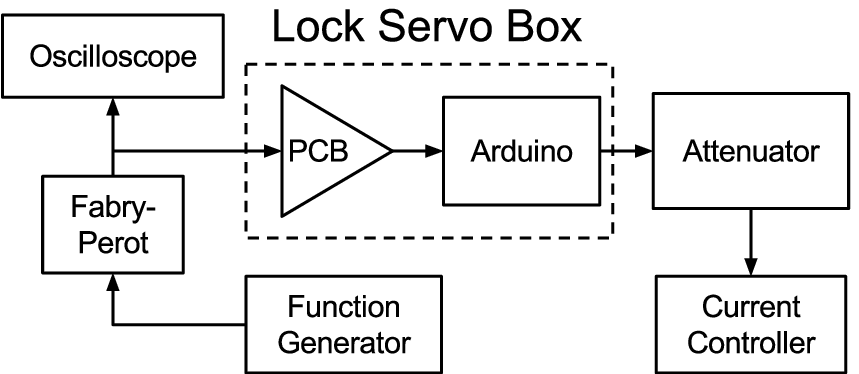}
                \label{fig:schematic}
                }
        \label{fig:setup}
        \caption{ (a) is a simplified optical layout of our injection lock setup, with laser beams marked in blue and electronic signals in black.  Function generators, Mirrors, $\lambda/2$ waveplates, polarizing beamsplitters, and mode-matching lenses are marked FG, M, HWP, PBS, and L respectively.  (b) is a schematic of the lock servo electronics.  The PCB consists of a unity-gain buffer amplifier and a summing amplifier to re-scale the FP signal for the microcontroller.}
\end{figure}

Our experimental setup, depicted in figure \subref*{fig:diagram}, is used for laser cooling applications with ytterbium atoms and is representative of most injection lock assemblies. The high power needed for atomic beam slowing is generated by injection locking a slave laser diode (LD, Mitsubishi Electric ML320G2-11, $120\,$mW) with a small portion (seed beam) of master light from an external cavity diode laser (ECDL, Toptica DL Pro). Under typical operating conditions the slave diode is held close to $100$mA with seed light power input at $300$$\mu$W and total slave power output at $80$mW.   The seed beam is spatially overlapped with the slave laser beam using plano-convex AR-coated cylindrical lenses and optically isolated using a custom-order  Faraday isolator (EOT $405$nm) optimized for $399$nm. The remainder of the master laser output is used to stabilize its center frequency by spectroscopy on the aforementioned  $\leftidx{^1}{S}{_0} \rightarrow \leftidx{^1}{P}{_1}$ transition, as well as for other laser cooling applications and diagnostics. A small portion of slave light is sent to a scanning Fabry-Perot (FP) cavity (Thorlabs SA200-3B) with free spectral range (FSR) $1.5\,$GHz and finesse 200.  The FP is controlled by an external function generator producing a triangle voltage sweep and FP transmission signals of both master and slave diodes are monitored. When the transmission signal of the slave laser resembles the narrow Lorentzian of the master, an injection lock is achieved. The FP transmission signal of the slave light is our main tool for monitoring the quality of an injection lock. When the slave diode falls out of an injection lock, the transmission signal jumps from a single narrow Lorentzian peak to several smaller, disordered peaks spread over a broader frequency range. The sensitivity of the FP transmission to the injection lock status of the slave diode allows us to automate both active injection lock maintenance and injection lock recovery by providing appropriate feedback to the slave diode current.
%
%

The current required for a stable injection lock can change over time as a result of drifts in the temperature of the slave diode and the alignment of the seed light. We will refer to the ``injection interval'' as the range of currents over which a stable injection lock is possible. If the injection interval changes so that it no longer includes the current which is being applied to the slave diode, then the injection lock is lost. The frequency of such events can be passively reduced by thermal and mechanical isolation of the setup. Here we demonstrate active control of the diode current to keep it within the injection interval by means of a device we will refer to as the lock servo.
 
Figure \subref*{fig:schematic} shows a schematic of the lock servo electronics.  While the design details are optimized to our application, they can easily be adjusted to other injection setups. The FP signal is sent through a unity gain buffer amplifier and a summing amplifier in order to prevent power draw on the Fabry-Perot, amplify the original waveform by a factor of 10, and add a DC offset to scale the waveform to within $0-3.3$V. The new signal is sent to an analog read pin on a microcontroller (Arduino Due A000062) with 12-bit precision, resulting in a resolution of $800 \mu$V/bit. The lock servo program takes this bit value as input and uses the DAC pin as output. The voltage level on the DAC pin of the microcontroller has 12-bit precision, but due to the design of the Arduino Due the output has a minimum and maximum of $1/6$ and $5/6$ of the internal logic levels, giving us $2^{12}$ divisions of the voltage range $.55$ to $2.75$ V. This output is sent through a 20 dB attenuator (Minicircuits HAT-20+) to the external current modulation port on the Thorlabs LDC 202C current controller for the slave diode. Given the modulation port's response of $20\,$mA/V, the lock servo has an effective current offset range of $4.4\,$mA and a resolution of $1.07\,\mu$A/bit over its full 12-bit output.

\section{\label{sec:injectionlocking}Injection Lock Characterization}

\begin{figure}
        \centering
        \subfloat[]{
				\includegraphics{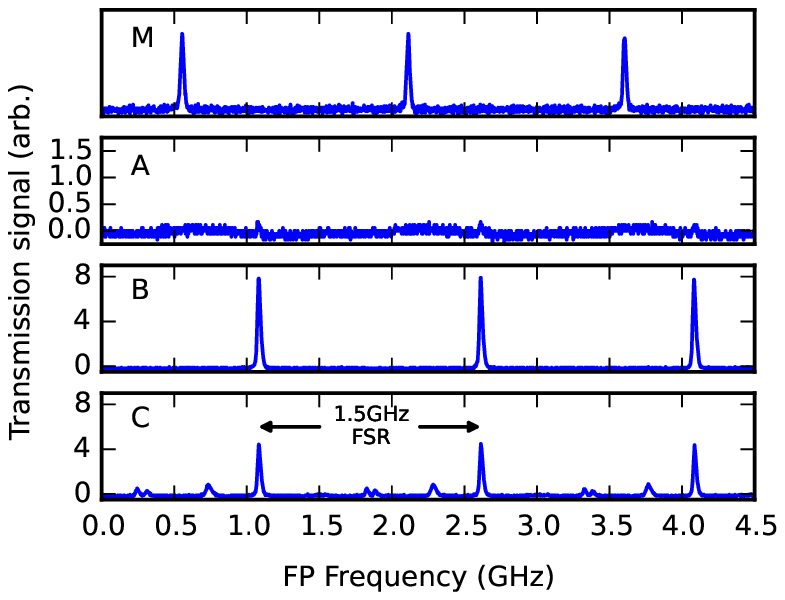}
                \label{fig:transmissionfigs}
                }
                
        \subfloat[]{
				\includegraphics{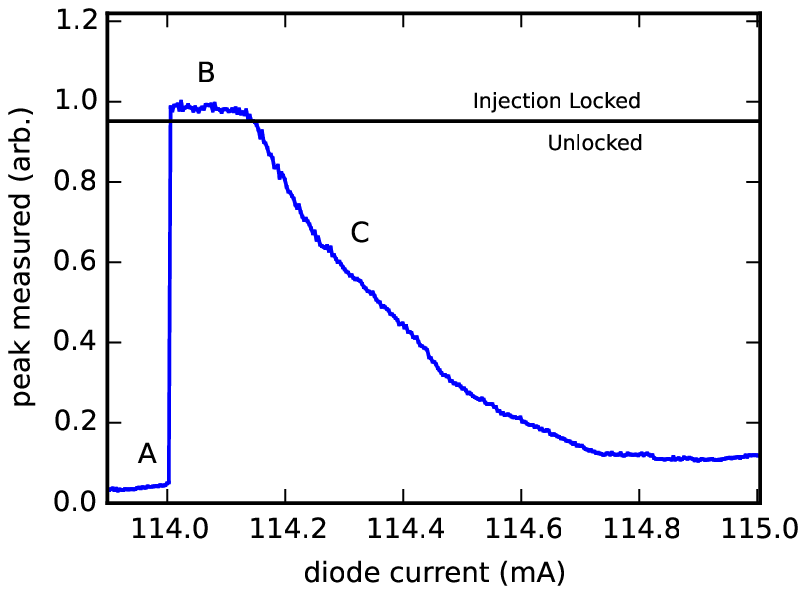}
                \label{fig:SpectralPurityCurve}
                }
        \caption{Assessment of the injection lock status using the Fabry-Perot transmission signal of the slave diode. For reference, the master laser FP signal (M) is depicted with frequency offset induced by an acousto-optic modulator.  The microcontroller reads the peak value of the FP transmission signal (a) of the slave laser while sweeping the diode current to produce spectral purity curves, with a representative example shown in (b). Intervals of injection lock (region B) are surrounded by regions of fully unlocked broad emission (region A) and partially injected multimode output (region C) when scanning the slave current. FP transmission signals of the slave for regions A, B, and C displayed in (a) correspond to current values $113.96, 114.10$, and $114.31\,$mA respectively. Injection intervals are separated by about $20\,$mA in diode current.}
        \label{scancurve}
\end{figure}

The typical response of the slave frequency spectrum to diode current near an injection interval is shown in Fig. \ref{scancurve}. The FP transmission signal (see Fig. \subref*{fig:transmissionfigs}) changes to a single sharp peak when the slave laser is injection locked. We characterize the variation in the quality of the injection lock with ``spectral purity curves" such as the one shown in Fig. \subref*{fig:SpectralPurityCurve}. This curve is a plot of the peak value of the FP transmission signal as a function of slave diode current. In typical operation the peak value is generated as the maximum FP output over a time interval of $200\,$ms, with the FP scan rate set to cover one FSR in less than $10\,$ms. The spectral purity curve is asymmetric: there is a gradual change in peak height on the high-current side and a rapid one on the low-current side. The two sides are separated by an approximately flat plateau region in the middle which corresponds to the useful injection interval. 

\begin{figure}
\centering
\includegraphics{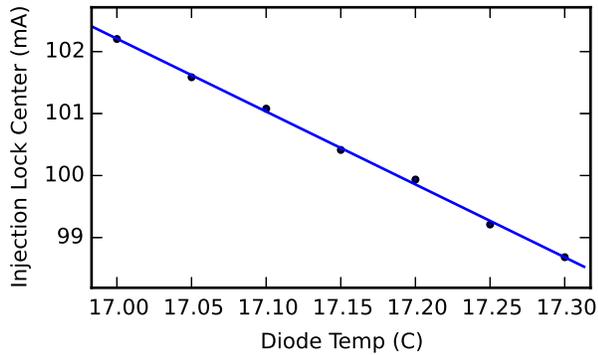}
\caption{Injection lock interval measurements while varying the temperature controller set point of the slave diode. The midpoint of the injection lock region for each temperature is plotted.  For the laser diode cavity to remain at resonance length for the seed light, thermal expansion due to diode heating must be offset by a decrease in thermal contribution from Joule heating.}
%
\label{fig:TempVariation}
\end{figure}

We conjecture that the asymmetry stems from the combination of Joule heating of the laser cavity from diode current \cite{wieman} and heating from seed light resonance. The current required for an injection lock strongly depends on the diode temperature as shown in Figure \ref{fig:TempVariation}.  At a diode current above the injection interval, the cavity length is larger than the resonance length from Joule heating. At a slightly smaller value of current, the Joule heating is reduced - shortening the cavity length. This is concurrent with the cavity length approaching resonance and the consequent higher level of intra-cavity seed light causes the cavity to warm and tend to lengthen. These two competing thermal effects thus act against each other to create a more gradual approach to injection lock on the high current side of the injection lock region. We note that our slave diode has only a generic AR coating. The strength of thermal effects from seed light resonance will depend on the specific AR coating of the diode. On the low current side of the injection interval, the two thermal effects act with each other instead of against, resulting in a rapid change of peak height with current. We have also observed a hysteresis in the shape of the spectral purity curve, indicated by a dependence of the width of the injection interval on the direction in which it is approached in current. This effect is also explained by the contribution of seed light to cavity expansion. When increasing diode current from a low value, there is no initial seed light to drive cavity expansion. Therefore an injection lock is not achieved until higher current values. As a result, the width of the injection interval is smaller when the curve is created with increasing current, and wider with decreasing current. The data in figure \subref*{fig:SpectralPurityCurve} is taken with decreasing current. In order to use maxima of the scanning FP signal to stabilize the injection lock we must scan over at least one FSR, which sets the bandwidth of any active feedback scheme. We observe the amplitude of the scanning FP signal to vary enough such that more traditional locking techniques (such as the side-of-fringe method) are not useful in our implementation. Our stabilization scheme relies only on the empirically observed shape of the spectral purity curve.

\section{\label{sec:lockservo}Injection Lock Servo}

We engage our lock servo after first manually obtaining the injection lock in the standard way. Once the lock servo is engaged, the FP transmission signal at that moment is used to define a threshold value by recording the maximum detected voltage over the past second. We define the threshold value to be 95\% of the maximum of the transmission signal over that time, and define an injection lock to occur when the peak FP signal is larger than the threshold value. Our choice for the threshold is guided by the signal-to-noise in the top flat part of the injection interval (see Fig. \subref*{fig:SpectralPurityCurve}) and the optimal threshold value can be different in other setups. The electronic feedback is implemented in two modes: recovery mode and active servo mode. When the transmission signal has a maximum above the threshold, the active servo program runs and when below the threshold, the recovery program runs. In recovery mode, it steps the offset current through a broader range of values in order to reacquire the injection lock.
The shape and hysteresis properties of the spectral purity curve mean that it is advantageous to approach the injection interval from above. Therefore, the first step of the recovery algorithm is to increase the diode current by $1$mA, which is enough to put the diode current above the injection interval. Next, it decreases the current in increments of the typical injection interval size of $.1$mA until the transmission peak detected is above threshold. Since the Arduino is outputting bit values in modular arithmetic, when a bit output of 0 is decremented the value shifts up by $2^{12}(= 4096)$, resulting in a looping scan of $4.4$mA around the injection lock region. As discussed below, this interval is wide enough to cover the largest lock drift from thermal effects we observe.

When the slave diode is injection locked, the active servo algorithm runs every 2.5 seconds. As noted earlier, the spectral purity curve is asymmetric. When the current is within the injection interval, lowering the current leads to an abrupt drop out of injection lock, whereas raising the current leads to a gradual decrease in Fabry-Perot peak height. The strategy of the active servo is to stay near the high current side of the injection interval plateau  (Fig. \subref*{fig:SpectralPurityCurve}) by detecting slight changes in the peak transmission. Starting from the existing current offset $i$ from the lock servo and some peak detected value of the transmission signal, the active follower program proceeds as follows: it samples the transmission signal for peaks at offset current values $i$ and $i \pm \delta$. If the peak detected at $i+\delta$ decreases by no more than $p_1$ percent, the new output is set to $i + \delta$. If the peak detected at $i - \delta$ increases by $p_2$ percent, the new output is set to $i-\delta$. To scan for each of the three peak values at $i$ and $i \pm \delta$ in the transmission signal, the Arduino compares the voltage read from the FP to a stored value and stores the higher of the two at a sampling rate of 267kHz over several FSR's. The ideal values for these variables depend on the shape and signal-to-noise properties of the spectral purity curve of the diode in use. For the work described here we have used the values $\delta = .01$mA, $p_1 = -0.25$, $p_2 = 0.5$.  This low-frequency feedback scheme has proven remarkably capable at tracking lock drifts in our setup. Other values, or even algorithms, can be used and are easily programmed into the microcontroller.

\section{\label{sec:performance}Device Performance}

\begin{figure}
\centering
\includegraphics{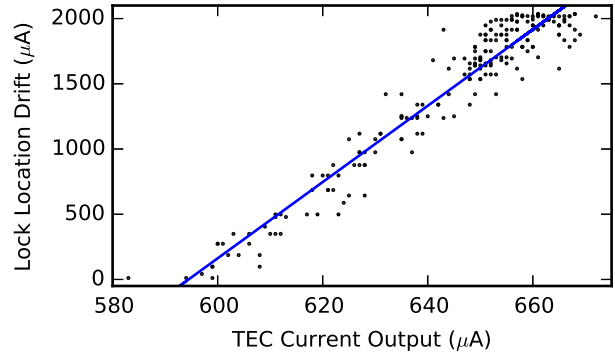}
\caption{Slave laser temperature controller current output over a two hour period after turning on slave laser current controller. The temperature controller was set to keep the laser diode at $17.06 ^\circ$C. As the temperature profile of the case approaches a steady state, the current needed for an injection lock changes. The data is fit well by a linear function (solid line).}
\label{tec}
\end{figure}

\begin{figure}
        \centering
        \subfloat[]{
                \includegraphics{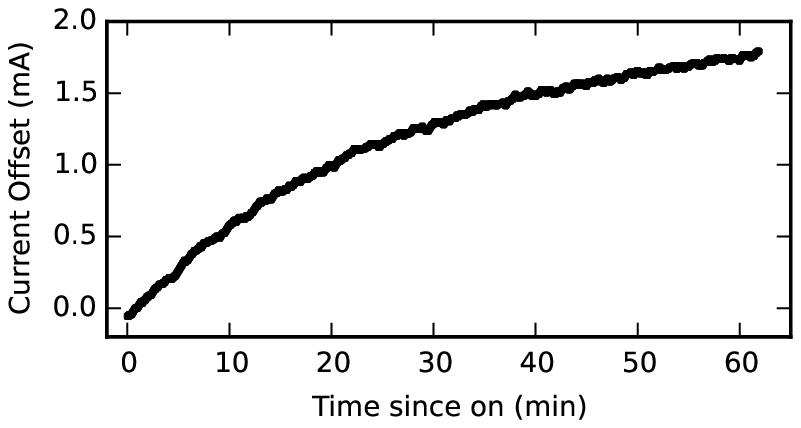}
                \label{fig:activefollower}
                }
        
        \subfloat[]{
                \includegraphics{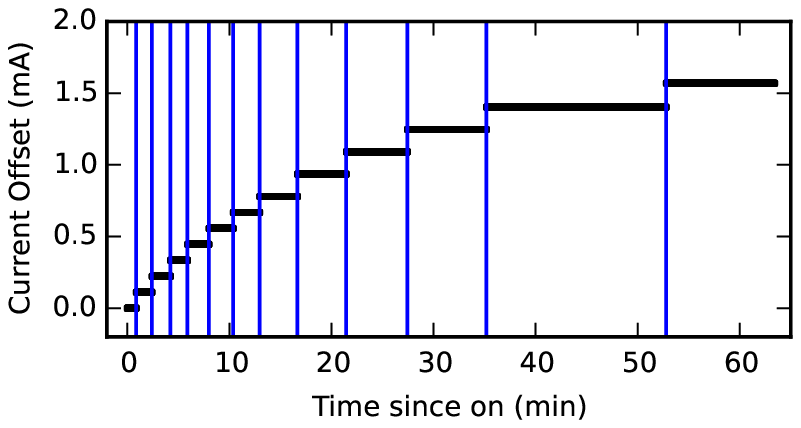}
                \label{fig:passivefollower}
                }
        \caption{Lock servo current offset output during the initial hour after turning on the slave laser current controller. (a) is with active servo program running, (b) is with only recovery program running. Vertical lines in (b) indicate when the injection lock was temporarily lost.}
        \label{hourdrift}
\end{figure}

We demonstrate the capabilities of our stabilization method by monitoring its performance starting immediately after turning on the slave diode current controller (figs. \ref{tec},\ref{hourdrift}), with all other experimental components in place beforehand. The largest movements of injection interval that we observe occur during this time while the slave diode is warming up. A Thorlabs TED 200C precision temperature controller is connected to a Thorlabs TCLDM9 diode case containing the high-power slave diode. Despite the temperature readout of the thermistor deviating by no more than $.01 ^\circ$C,  the injection interval is measured to drift by $2$mA over the course of two hours after being turned on. This is due to the cavity length's strong dependence on overall temperature profile and not just the temperature measured at the thermistor. Figure \ref{tec} is a plot of the measured lock location drift in current against the current output of the temperature controller driving the thermoelectric cooler (TEC) inside the diode case. A linear fit to this data shows that the ideal injection lock current drifts by $30\,\mu$A for every $1\,\mu$A change in TEC current. Since the interval over which injection locks are stable in our setup typically range from 100 to 200$\,\mu$A, this illustrates the strong temperature sensitivity of the ideal diode current value compared to the injection interval for stable locking.

Figure \ref{hourdrift} demonstrates the individual performance of the active servo (Fig. \subref*{fig:activefollower}) and recovery (Fig. \subref*{fig:passivefollower}) programs operating for one hour while this initial drift occurs. The active servo successfully tracks the lock drift over the entire hour while the recovery algorithm must re-lock whenever the spectral purity curve drifts further than the width of the injection interval. The offset current output from the active servo and recovery mode programs are shown in Figs. \subref*{fig:activefollower} and \subref*{fig:passivefollower} respectively. Note that without the lock servo, a user would need to manually adjust the diode current to the slave laser at every vertical line in Fig. \subref*{fig:passivefollower} to maintain an injection lock.

\begin{figure}
\centering
\includegraphics{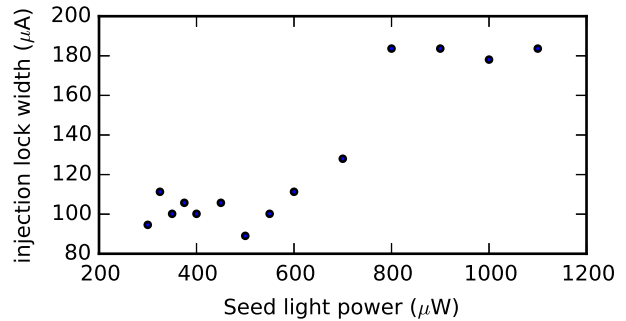}
\caption{Injection interval width of peak transmission above threshold for various seed light intensities. The lock servo tracks the decreasing edge of the spectral purity response and does not require a large interval of injection, reducing the need
for large seed light intensities.}
\label{fig:SeedLightPower}
\end{figure}

The use of the lock servo also allows for lower values of seed light power for injection locking. Our measurements of the spectral purity curve for various seed light powers show that the injection interval increases as seed light power increases (Fig. \ref{fig:SeedLightPower}). Using high-power increases passive stability since the injection interval position would need to drift over a larger distance in current before the injection lock is lost. The lock servo tracks the high-current side of the injection interval plateau, even if the injection interval is small. Only $300\mu$W seed light power was used for the data shown in Figs. \ref{tec} and \ref{hourdrift}. In our setup, this seed power produces an injection interval which is a factor of two less than that at $1\,$mW seed power, where the passive stability is much greater. These observations show that the lock servo allows operation with seed power level at least a factor of three below that needed for optimum passive stability. This is a useful feature for optical setups where master laser power is at a premium.   

Injection locks can be notoriously unstable - mechanical vibrations of the master or slave laser, mode hops in the master, or accidental beam blocking can all cause the slave laser to fall out of injection. There is often no guarantee the injection lock will recover when the disturbance is gone since the current needed for injection locking usually changes. However, the new diode current needed to re-lock the slave laser is invariably close to the old value. When these disturbances happen the transmission signal decreases below threshold and the recovery follower program runs, continually sweeping across the full $4.4$mA output until the injection lock is recovered. In practice this is relatively quick - each vertical line on figure \ref{hourdrift} represents an unlocking of the slave laser, which the recovery follower re-locked after an average time of 2 seconds. The same automatic recovery happens for mechanical disturbances, blocked beams, or transient issues with a master laser, once the disturbance is removed.

\section{\label{sec:conclusion}Conclusion}

In summary, we have developed a useful addition to the laser frequency stabilization toolkit for atomic and optical physics applications: a device which can actively maintain an injection lock on a slave laser diode. Our lock servo device is relatively straightforward to implement and provides the additional benefits of less optical power requirements for the seed light and alleviation of stringent passive stability needs on the optical setup. All of these features have significantly improved the stability and reliability of our injection lock systems for Yb laser cooling. Our results can directly help any experimental effort with injection locked lasers for cooling or addressing Yb, an atom which has various applications in atomic clocks \cite{hinkley}, preparation of quantum degenerate systems \cite{takasu,hansen}, precision atom interferometry \cite{jamison}, quantum simulation \cite{pagano, scazza} and quantum information processing \cite{stock}. We expect our results to also be applicable to diode lasers operating at other wavelengths and therefore to be of general use for a broad range of diode laser injection lock setups and applications. 

\begin{acknowledgments}
We acknowledge support from the National Science Foundation.  The authors thank R. Weh, K. McAlpine, D. Gochnauer and A. Jamison for technical help and useful discussions.
\end{acknowledgments}

%

\end{document}